\documentclass[prd,aps,nofootinbib,floatfix,10pt]{revtex4}

\usepackage{amsmath,graphicx,epsfig,amssymb,dsfont,mathtools}
\usepackage[usenames]{color}
\usepackage{ulem} 
\usepackage{bigstrut}
\usepackage{slashed}
\usepackage{gensymb}
\usepackage{multirow}
\usepackage{subfigure}

\allowdisplaybreaks

\begin{document}
	
\title{Revisiting $\Xi_{Q}-\Xi_{Q}^{\prime}$ mixing in QCD sum rules}
	
\author{
Xiao-Yu Sun$^{1}$, 
Fu-Wei Zhang$^{1}$, 
Yu-Ji Shi$^{2}$~\footnote{Email:shiyuji@ecust.edu.cn},
Zhen-Xing Zhao$^{1}$~\footnote{Email:zhaozx19@imu.edu.cn}
}

\affiliation{
$^{1}$ School of Physical Science and Technology, Inner Mongolia University,
Hohhot 010021, China, \\
$^{2}$ School of Physics, East China University of Science and Technology,
Shanghai 200237, China
}
		
\begin{abstract}
In this work, we perform a QCD sum rules analysis on the $\Xi_{Q}-\Xi_{Q}^{\prime}$
mixing. Contributions from up to dimension-6 four-quark operators 
are considered. However, it turns out that, only dimension-4 and dimension-5 operators
contribute, which reveals the non-perturbative nature of mixing. Especially
we notice that only the diagrams with the two light quarks participating
in gluon exchange contribute to the mixing. Our results indicate that
the mixing angle $\theta_{c}=(1.2\sim2.8)^{\circ}$ for the $Q=c$ case and 
$\theta_{b}=(0.28\sim0.34)^{\circ}$ for the $Q=b$ case. 
Our prediction of $\theta_{c}$ is consistent with
the most recent Lattice QCD result within error. Such a small mixing
angle seems unlikely to resolve the tension between the recent experimental
measurement from Belle and Lattice QCD calculation for the
semileptonic decay $\Xi_{c}^{0}\to\Xi^{-} e^{+}\nu_{e}$.  
\end{abstract}

\maketitle

\section{Introduction}

The semileptonic decays of hadrons are of great significance for extracting
CKM matrix elements and testing the standard model. Recently, the
semileptonic decay $\Xi_{c}^{0}\to\Xi^{-} e^{+}\nu_{e}$ is measured by the
Belle collaboration~\cite{Belle:2021crz}
\begin{equation}
{\cal B}(\Xi_{c}^{0}\to\Xi^{-}e^{+}\nu_{e})=(1.31\pm0.39)\%,
\end{equation}
while the Lattice QCD prediction in Ref.~\cite{Zhang:2021oja} is 
\begin{equation}
{\cal B}(\Xi_{c}^{0}\to\Xi^{-}e^{+}\nu_{e})=(2.38\pm0.44)\%.
\end{equation}
Our preliminary calculation based on QCD sum rules in Ref.~\cite{Zhao:2021sje} even
gives a larger result 
\begin{equation}
{\cal B}(\Xi_{c}^{0}\to\Xi^{-}e^{+}\nu_{e})=(3.4\pm0.7)\%.
\end{equation}
It can be seen that there exist one tension between experimental data
and theoretical predictions. 

In Refs.~\cite{He:2021qnc,Geng:2022xfz,Geng:2022yxb,Ke:2022gxm}, the authors suggested that this puzzle
can be resolved by considering $\Xi_{c}-\Xi_{c}^{\prime}$ mixing
on theoretical side. If so, one would expect that there exists
a sizable $\Xi_{c}-\Xi_{c}^{\prime}$ mixing angle. Some efforts have
been made in this direction. Early in 2010, a QCD sum rules analysis
was performed, and the authors arrived at $\theta_{c}=5.5\degree\pm1.8\degree$~\cite{Aliev:2010ra}.
In Ref.~\cite{Matsui:2020wcc}, this mixing angle is obtained
as $|\theta_{c}|=8.12\degree\pm0.80\degree$ in heavy quark effective
theory. In Ref.~\cite{Liu:2023feb}, the result of Lattice QCD shows
that this mixing angle is equal to $1.2\degree\pm0.1\degree$. More
theoretical predictions can be found in Table \ref{Tab:comparison}. 

One can see that, large differences exist among different theoretical
predictions. In this work, we intend to perform a new QCD sum rules
analysis. First of all, it is necessary to figure out the concepts
of flavor eigenstates and mass eigenstates. The flavor eigenstates
are defined as follows 
\begin{align}
\Xi_{Q}^{\bar{3}} & =\frac{1}{\sqrt{2}}(qs-sq)Q,\nonumber \\
\Xi_{Q}^{6} & =\frac{1}{\sqrt{2}}(qs+sq)Q\label{eq:flavor_eigenstates}
\end{align}
with $Q=c,b$ and $q=u,d$. Eqs. (\ref{eq:flavor_eigenstates}) are
of course the classification of quark model, where $\Xi_{Q}^{\bar{3}}$
belongs to the SU(3) flavor antitriplet, and $\Xi_{Q}^{6}$ belongs
to the sextet, as indicated by their notations. The two light quarks
are usually considered to form a scalar diquark and an axial-vector
diquark in $\Xi_{Q}^{\bar{3}}$ and $\Xi_{Q}^{6}$, respectively.
In reality, the physical mass eigenstates $\Xi_{Q}$ and $\Xi_{Q}^{\prime}$
are the mixing of flavor eigenstates 
\begin{equation}
\left(\begin{array}{c}
|\Xi_{Q}\rangle\\
|\Xi_{Q}^{\prime}\rangle
\end{array}\right)=\left(\begin{array}{cc}
\cos\theta & \sin\theta\\
-\sin\theta & \cos\theta
\end{array}\right)\left(\begin{array}{c}
|\Xi_{Q}^{\bar{3}}\rangle\\
|\Xi_{Q}^{6}\rangle
\end{array}\right).\label{eq:mixing}
\end{equation}

Although there already exists a QCD sum rules analysis in Ref.~\cite{Aliev:2010ra},
while in this work, we will highlight the following points:
\begin{itemize}
\item New definitions (see Eq. (\ref{eq:interpolating_currents})) of interpolating
currents are adopted. These definitions have been proved in a quark
model~\cite{Zhao:2023yuk}, and are considered to be possibly better
definitions of interpolating currents for baryons. 
\item We attempt to reveal the nature of mixing. Through detailed
calculation, one can clearly see that the gluon exchange involving the two light quarks
plays a crucial role in flavor mixing. It is gluon exchange
that can change the spin of the system of two light quarks.
\item In the heavy quark limit, the spin of the system of two light quarks
is a good quantum number, therefore, the mixing angle between $\Xi_{Q}$
and $\Xi_{Q}^{\prime}$ should be zero. Our calculation results show
such a trend. 
\end{itemize}

The rest of this article is arranged as follows. In Sec. II, QCD sum
rules analysis is performed, contributions from up to dimension-6 four-quark operators
are considered. In Sec. III, numerical results are shown and are
compared with other predictions in the literature. We conclude this article
in the lat section. 

\section{QCD sum rules analysis }

The mass sum rule for $\Xi_{Q}^{(\prime)}$ can be obtained by considering
the following two-point correlation function 
\begin{equation}
\Pi^{(\prime)}(p)=i\int d^{4}xe^{ip\cdot x}\langle0|T\{J^{(\prime)}(x)\bar{J}^{(\prime)}(0)\}|0\rangle,\label{eq:cf_XiQ_XiQp}
\end{equation}
where $J^{(\prime)}$ stands for the interpolating current of the mass
eigenstate $\Xi_{Q}^{(\prime)}$. It is natural to expect that $\bar{J}^{(\prime)}$
creates only $\Xi_{Q}^{(\prime)}$ and not the other one, and in this
sense, the following two correlation functions should be zero 
\begin{align}
i\int d^{4}xe^{ip\cdot x}\langle0|T\{J(x)\bar{J}^{\prime}(0)\}|0\rangle & =0,\nonumber \\
i\int d^{4}xe^{ip\cdot x}\langle0|T\{J^{\prime}(x)\bar{J}(0)\}|0\rangle & =0.\label{eq:cf_to_get_theta}
\end{align}
$J$ and $J^{\prime}$ are linear combinations of $J_{0}$ and $J_{1}$ --
the interpolating currents of flavor eigenstates $\Xi_{Q}^{\bar{3}}$
and $\Xi_{Q}^{6}$:
\begin{equation}
\left(\begin{array}{c}
J\\
J^{\prime}
\end{array}\right)=\left(\begin{array}{cc}
\cos\theta & \sin\theta\\
-\sin\theta & \cos\theta
\end{array}\right)\left(\begin{array}{c}
J_{0}\\
J_{1}
\end{array}\right),\label{eq:currents_mixing}
\end{equation}
which is a counterpart of Eq. (\ref{eq:mixing}). However, it should be noted that
since there is no exact one-to-one correspondence between the interpolating current and the hadron state, 
the quark-hadron duality ansatz is actually implicit in Eq. (\ref{eq:currents_mixing}).
In this work, $J_{0,1}$ are given by
\begin{align}
J_{0} & =\epsilon_{abc}[q_{a}^{T}C\gamma_{5}(1+\slashed v)s_{b}]Q_{c},\nonumber \\
J_{1} & =\epsilon_{abc}[q_{a}^{T}C(\gamma^{\mu}-v^{\mu})(1+\slashed v)s_{b}]\frac{1}{\sqrt{3}}\gamma_{\mu}\gamma_{5}Q_{c},\label{eq:interpolating_currents}
\end{align}
where $a,b,c$ are color indices, and $v^{\mu}\equiv p^{\mu}/\sqrt{p^{2}}$
is the 4-velocity of baryon. As mentioned in the Introduction, these
new definitions have been proved in a quark model, and are possibly
better definitions of interpolating currents for baryons. 

It can be seen from Eqs. (\ref{eq:cf_XiQ_XiQp}) and (\ref{eq:cf_to_get_theta})
that, we need to calculate the following 4 correlation functions 
\begin{equation}
\Pi_{ij}(p)=i\int d^{4}xe^{ip\cdot x}\langle0|T\{J_{i}(x)\bar{J}_{j}(0)\}|0\rangle\label{eq:four_cfs}
\end{equation}
with $i,j=0,1$. 

From Eq. (\ref{eq:cf_XiQ_XiQp}), one can obtain the mass sum rules
for $\Xi_{Q}$ and $\Xi_{Q}^{\prime}$ 
\begin{align}
\Pi=\Pi_{00}\cos^{2}\theta+\Pi_{11}\sin^{2}\theta+\Pi_{01}\sin2\theta & ,\label{eq:sum_rule_XiQ}\\
\Pi^{\prime}=\Pi_{11}\cos^{2}\theta+\Pi_{00}\sin^{2}\theta-\Pi_{01}\sin2\theta & .\label{eq:sum_rule_XiQp}
\end{align}
As explicit calculation has shown, $\Pi_{01}=\Pi_{10}$, then from
Eq. (\ref{eq:cf_to_get_theta}), one can arrive at 
\begin{equation}
\Pi_{01}\cos2\theta+(\Pi_{11}-\Pi_{00})\frac{1}{2}\sin2\theta=0,
\end{equation}
or 
\begin{equation}
\tan2\theta=\frac{2\ \Pi_{01}}{\Pi_{00}-\Pi_{11}}.\label{eq:mixing_angle_formula}
\end{equation}

One can easily check that the above description is equivalent to the
following matrix diagonalization formula
\begin{equation}
O\Pi O^{-1}=\Pi_{{\rm diag}}
\end{equation}
with 
\begin{equation}
\Pi=\left(\begin{array}{cc}
\Pi_{00} & \Pi_{01}\\
\Pi_{01} & \Pi_{11}
\end{array}\right),\quad O=\left(\begin{array}{cc}
\cos\theta & \sin\theta\\
-\sin\theta & \cos\theta
\end{array}\right),\quad\Pi_{{\rm diag}}=\left(\begin{array}{cc}
\Pi & 0\\
0 & \Pi^{\prime}
\end{array}\right).
\end{equation}

One important note. From Eq. (\ref{eq:mixing_angle_formula}), one
can see that, we had better normalize the two interpolating currents
in Eq. (\ref{eq:interpolating_currents}) to a same
factor, and we have indeed done that. Therefore, in this work, $\Pi_{00}$,
$\Pi_{11}$, and $\Pi_{01}$ are on an equal footing, and we can explicitly
compare their respective contributions from the same dimensions at the
QCD level, see below. 

In this work, we calculate the 4 correlation functions in Eq. (\ref{eq:four_cfs}),
considering the contributions from perturbative term (dim-0), quark
condensate (dim-3), gluon condensate (dim-4), quark-gluon condensate
(dim-5), and four-quark condensate (dim-6), as can be seen in Fig.~\ref{fig:dim03456}.
The analytical results are listed in Appendix A.
Through detailed calculation, one can clearly see that: 
\begin{itemize}
\item For $\Pi_{01}$, it turns out that, only 4 diagrams are nonzero--they
are dim-4(a,b) and dim-5(a,c). The physical meaning of $\Pi_{01}$
is: it provides the absolute possibility for the diquark to transition
from $0^{+}$ to $1^{+}$, or vice versa. As far as we are concerned,
the mixing between $\Xi_{Q}^{\bar{3}}$ and $\Xi_{Q}^{6}$ originates
from that the two light quarks exchange gluons with the background
fields in vacuum, and with the heavy quark $Q$. 
\item For $\Pi_{00}$ and $\Pi_{11}$, dim-0,3,6, and dim-4(d,e,f) are respectively
equal to each other, so they do not contribute to the denominator
$\Pi_{00}-\Pi_{11}$ in Eq. (\ref{eq:mixing_angle_formula}). Only
dim-4(a,b,c) and dim-5(a,b,c,d) contribute to $\Pi_{00}-\Pi_{11}$.
The physical meaning of $\Pi_{00}-\Pi_{11}$ is: it measures the difference,
or the ``gap'' between $\Xi_{Q}^{\bar{3}}$
and $\Xi_{Q}^{6}$; The larger the difference, the less likely the
two flavor eigenstates are to mix. 
\end{itemize}

\begin{figure}[!]
\includegraphics[width=1.0\columnwidth]{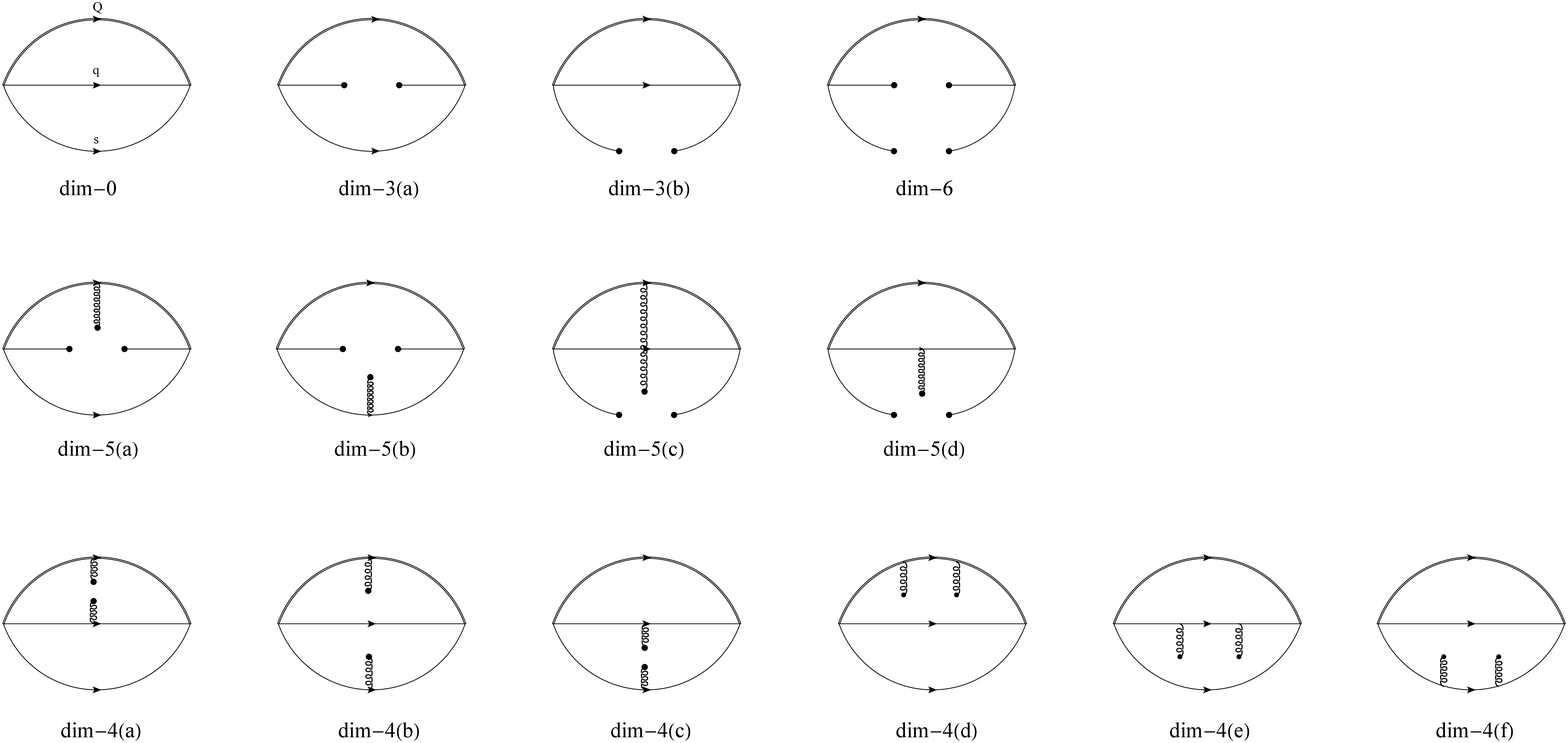}
\caption{All the diagrams considered in this work. We calculate all these diagrams for the 4 correlation functions $\Pi_{ij}$ with $i,j=0,1$. }
\label{fig:dim03456}
\end{figure}

The mixing angle formula in Eq. (\ref{eq:mixing_angle_formula}) is
of course our main research object. However, the corresponding QCD sum rules
are very different from the traditional
ones: it does not have the hadron-level representation. For this point,
try to consider the hadron-level representation of $\Pi_{01}$. That
is, Eq. (\ref{eq:mixing_angle_formula}) has only a representation
at the QCD level. The continuum threshold parameter $\sqrt{s_{0}}$
and Borel parameters $T^{2}$ cannot determined by the methods commonly
used in the literature. However, note that a reasonable threshold
parameter for Eq. (\ref{eq:mixing_angle_formula}) should lie between
those of $\Xi_{Q}$ and $\Xi_{Q}^{\prime}$. Naturally, in the following,
we present the mass sum rule of $\Xi_{Q}^{(\prime)}$. 

\subsection{The mass sum rule}

Since our preliminary results indicate that $\theta_{c}$ and $\theta_{b}$
are very small, Eqs. (\ref{eq:sum_rule_XiQ}) and (\ref{eq:sum_rule_XiQp})
are reduced into 
\begin{align}
\Pi=\Pi_{00} & ,\label{eq:sum_rule_XiQ_reduced}\\
\Pi^{\prime}=\Pi_{11} & .\label{eq:sum_rule_XiQp_reduced}
\end{align}
Following the same steps in Refs.~\cite{Zhao:2020mod,Zhao:2021sje},
one can perform QCD sum rules analysis on the correlation functions
$\Pi_{00,11}$ as follows. 

At the hadron level, after inserting the complete set of hadronic
states, one can obtain
\begin{equation}
\Pi^{{\rm had}}(p)=\lambda_{+}^{2}\frac{\slashed p+M_{+}}{M_{+}^{2}-p^{2}}+\lambda_{-}^{2}\frac{\slashed p-M_{-}}{M_{-}^{2}-p^{2}}+\cdots,
\end{equation}
where $\lambda_{+(-)}$ and $M_{+(-)}$ are respectively the pole
residue and mass of the positive-parity (negative-parity) baryon.
The pole residues of positive-parity and negative-parity baryons are
respectively defined by
\begin{align}
\langle0|J_{+}|{\cal B}_{+}(p,s)\rangle & =\lambda_{+}u(p,s),\nonumber \\
\langle0|J_{+}|{\cal B}_{-}(p,s)\rangle & =\lambda_{-}(i\gamma_{5})u(p,s).
\end{align}

At the QCD level, the correlation function is also calculated. In
this work, contributions from up to dimension-6 four quark operators are considered,
as can be seen in Fig. \ref{fig:dim03456}. The corresponding results
can be formally rewritten as 
\begin{equation}
\Pi^{{\rm QCD}}(p)=A(p^{2})\slashed p+B(p^{2}).
\end{equation}
The coefficient functions $A(p^{2})$ and $B(p^{2})$ are further
written into dispersion relations 
\begin{equation}
A(p^{2})=\int ds\frac{\rho^{A}(s)}{s-p^{2}},\quad B(p^{2})=\int ds\frac{\rho^{B}(s)}{s-p^{2}}.
\end{equation}

Using the quark-hadron duality assumption, and after performing the
Borel transformation, one can arrive at the following sum rule for
the positive-parity baryon
\begin{equation}
(M_{+}+M_{-})\lambda_{+}^{2}e^{-M_{+}^{2}/T_{+}^{2}}=\int^{s_{+}}ds(M_{-}\rho^{A}(s)+\rho^{B}(s))e^{-s/T_{+}^{2}},\label{eq:mass_sum_rule}
\end{equation}
where $s_{+}$ and $T_{+}^{2}$ are respectively the continuum threshold
parameter and Borel parameter. From Eq. (\ref{eq:mass_sum_rule}),
one can obtain the mass of the $1/2^{+}$ baryon
\begin{equation}
M_{+}^{2}=\frac{\int^{s_{+}}ds(M_{-}\rho^{A}+\rho^{B})\ s\ e^{-s/T_{+}^{2}}}{\int^{s_{+}}ds(M_{-}\rho^{A}+\rho^{B})\ e^{-s/T_{+}^{2}}}.\label{eq:mass_formula}
\end{equation}
In practice, Eq. (\ref{eq:mass_formula}) can be viewed as a constraint of Eq.
(\ref{eq:mass_sum_rule}), in which $M_{+}$ is required to be equal
to the experimental value of the positive-parity baryon. In this way, the threshold
parameter can be determined.

\section{Numerical results}

The following parameters are adopted~\cite{ParticleDataGroup:2022pth}:
\begin{align}
 & m_{c}(m_{c})=1.27\pm0.02\ {\rm GeV},\quad m_{s}(2\ {\rm GeV})=0.093\pm0.009\ {\rm GeV},\nonumber\\
 & m_{b}(m_{b})=4.18\pm0.03\ {\rm GeV}.
\end{align}
The condensate parameters are taken as~\cite{Colangelo:2000dp}: $\langle\bar{q}q\rangle(1\ {\rm GeV})=-(0.24\pm0.01\ {\rm GeV})^{3}$,
$\langle\bar{s}s\rangle=(0.8\pm0.2)\langle\bar{q}q\rangle$, and $\langle g_{s}^{2}G^{2}\rangle=(0.47\pm0.14)\ {\rm GeV}^{4}$,
and $\langle\bar{q}g_{s}\sigma Gq\rangle=m_{0}^{2}\langle\bar{q}q\rangle$
and $\langle\bar{s}g_{s}\sigma Gs\rangle=m_{0}^{2}\langle\bar{s}s\rangle$
with $m_{0}^{2}=(0.8\pm0.2)\ {\rm GeV}^{2}$. The renormalization
scale is taken as $\mu_{c}=1\sim3\ {\rm {\rm GeV}}$, and $\mu_{b}=3\sim6\ {\rm {\rm GeV}}$,
from which, one can estimate the dependence of calculation results
on the energy scale. 

\begin{figure}[!]
\includegraphics[width=1.0\columnwidth]{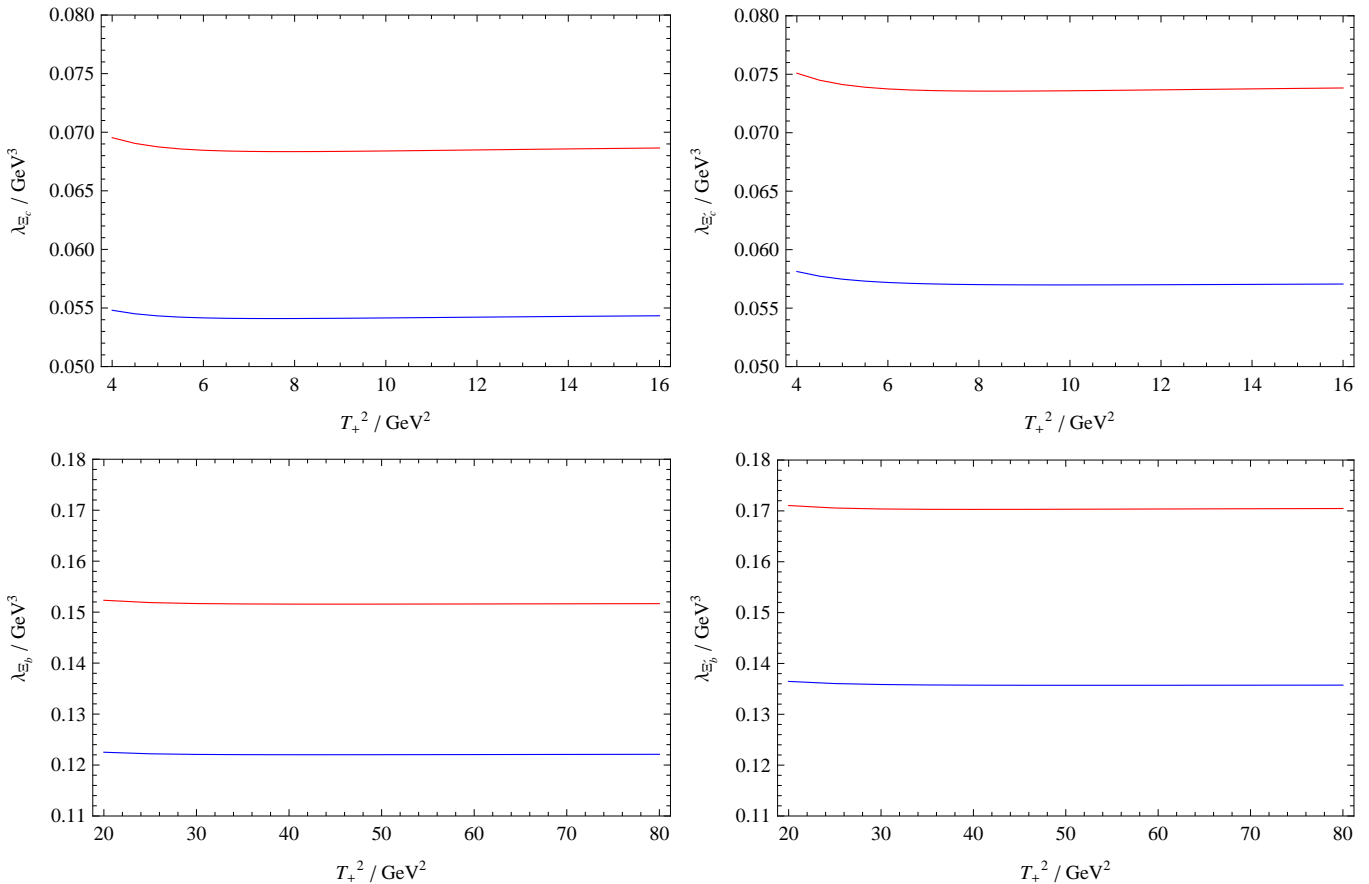}
\caption{
Pole residues of $\Xi_{Q}^{(\prime)}$ with $Q=c,b$. The blue lines correspond to the energy scale $\mu=m_{Q}$,
while the red lines correspond to the energy scale $\mu=3\ {\rm GeV}$ for $\Xi_{c}^{(\prime)}$
and $\mu=6\ {\rm GeV}$ for $\Xi_{b}^{(\prime)}$. The selections of $\sqrt{s_{0}}$ can be found in Table \ref{Tab:optimal_params_XiQ_XiQp}.
}
\label{fig:pole_residues}
\end{figure}

Following similar steps in Refs.~\cite{Zhao:2020mod,Zhao:2021sje}, one can
arrive at the optimal parameter selections for continuum thresholds $\sqrt{s_{0}}$
and Borel parameters $T^{2}$ for $\Xi_{Q}$ and $\Xi_{Q}^{\prime}$.
The corresponding results can be found in Fig. \ref{fig:pole_residues} and Table \ref{Tab:optimal_params_XiQ_XiQp}.
Some comments are given in order.
\begin{itemize}
\item As expected in Ref. \cite{Zhao:2023yuk}, the pole residues of $\Xi_{Q}$
and $\Xi_{Q}^{\prime}$ are almost equal when the interpolating currents
in Eq. (\ref{eq:interpolating_currents}) are used. 
\item The continuum threshold and Borel
parameter at the minimum point in Fig. \ref{fig:pole_residues} are selected
as the optimal parameters, as can be seen in Table \ref{Tab:optimal_params_XiQ_XiQp}.
These optimal parameters correspond to the experimental value of the baryon mass. 
\item As can be seen in Table \ref{Tab:optimal_params_XiQ_XiQp}, the optimal parameter selection satisfies:
$\sqrt{s_{0}}$ is about 0.5 GeV higher than the corresponding baryon
mass, and $T^{2}\sim{\cal O}(m_{H}^{2})$ with $m_{H}$ the baryon mass. 
\item As can be seen in Fig. \ref{fig:pole_residues}, the dependence of pole residues on the Borel parameters is weak, while
they are sensitive to changes in energy scales. The latter leads to
the main source of error. 
\end{itemize}

\begin{table}
\caption{Optimal parameter selections for the continuum thresholds $\sqrt{s_{0}}$
and Borel parameters $T^{2}$ for $\Xi_{Q}$ and $\Xi_{Q}^{\prime}$,
with $Q=c,b$. The central values are obtained at $\mu=m_{c}$ for
$\Xi_{c}^{(\prime)}$ and $\mu=m_{b}$ for $\Xi_{b}^{(\prime)}$.
The masses of $\Xi_{c}^{(\prime)0}(csd)$
and $\Xi_{b}^{(\prime)-}(bsd)$ are also listed as references \cite{ParticleDataGroup:2022pth}.}
\label{Tab:optimal_params_XiQ_XiQp}
\begin{tabular}{c|c|c|c}
\hline 
 & $\sqrt{s_{0}}/{\rm GeV}$  & $T^{2}/{\rm GeV}^{2}$ & Mass/GeV\tabularnewline
\hline 
$\Xi_{c}$  & for $\mu=m_{c}$, $2.95$; for $\mu=3\ {\rm GeV}$, $3.00$ & $\approx8$ & $2.470$\tabularnewline
$\Xi_{c}^{\prime}$  & for $\mu=m_{c}$, $3.02$; for $\mu=3\ {\rm GeV}$, $3.10$ & $10\pm2$ & $2.579$\tabularnewline
$\Xi_{b}$  & for $\mu=m_{b}$, $6.27$; for $\mu=6\ {\rm GeV}$, $6.30$  & $\approx40$ & $5.797$\tabularnewline
$\Xi_{b}^{\prime}$  & for $\mu=m_{b}$, $6.40$; for $\mu=6\ {\rm GeV}$, $6.45$  & $50\pm10$ & $5.935$\tabularnewline
\hline 
\end{tabular}
\end{table}

\begin{figure}[!]
\includegraphics[width=1.0\columnwidth]{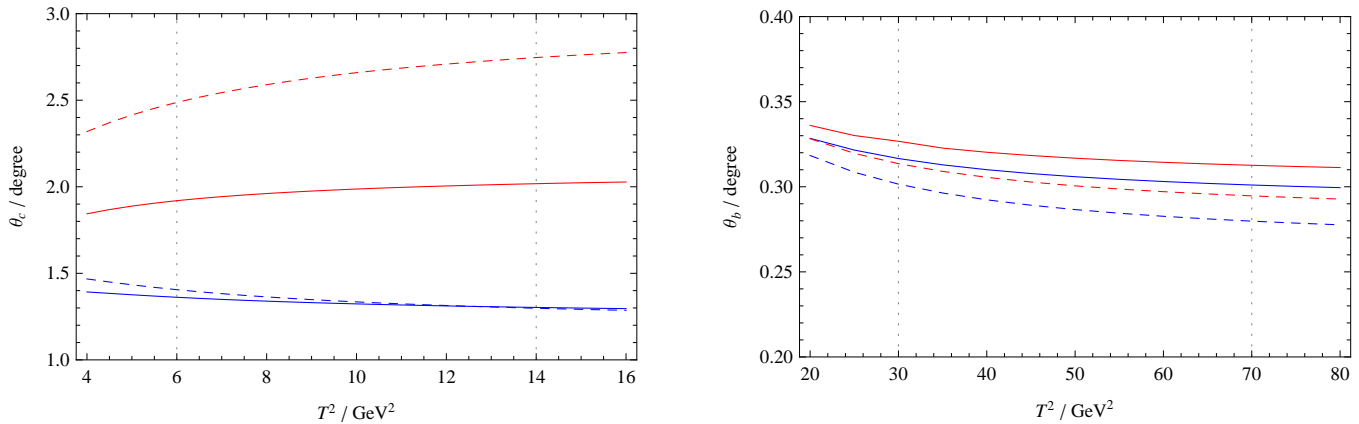}
\caption{Our predictions for $\theta_{c}$ and $\theta_{b}$. For the left figure, the solid blue and solid red lines respectively represent the curves of $\theta_{c}$ obtained from the first and second sum rules, where $\mu=m_{c}$, $\sqrt{s_{0}}=2.98\ {\rm GeV}$; The blue dashed and red dashed lines respectively represent the curves of $\theta_{c}$ obtained from the first and second sum rules, where $\mu=3\ {\rm GeV}$, $\sqrt{s_{0}}=3.05\ {\rm GeV}$. For the right figure, the solid blue and solid red lines respectively represent the curves of $\theta_{b}$ obtained from the first and second sum rules, where $\mu=m_{b}$, $\sqrt{s_{0}}=6.33\ {\rm GeV}$; The blue dashed and red dashed lines respectively represent the curves of $\theta_{b}$ obtained from the first and second sum rules, where $\mu=6\ {\rm GeV}$, $\sqrt{s_{0}}=6.38\ {\rm GeV}$. }
\label{fig:theta_c_and_theta_b}
\end{figure}

For the sum rule in Eq. (\ref{eq:mixing_angle_formula}), considering the continuum threshold should
lie between those of $\Xi_{Q}$ and $\Xi_{Q}^{\prime}$, and assuming $T^{2}\sim{\cal O}(m_{H}^{2})$, we choose
the following parameters:
\begin{itemize}
\item For $\theta_{c}$, when $\mu=m_{c}$, $\sqrt{s_{0}}=2.98\ {\rm GeV}$,
and when $\mu=3\ {\rm GeV}$, $\sqrt{s_{0}}=3.05\ {\rm GeV}$; the
Borel parameters $T^{2}\in[6,14]\ {\rm GeV}^{2}$.
\item For $\theta_{b}$, when $\mu=m_{b}$, $\sqrt{s_{0}}=6.33\ {\rm GeV}$,
and when $\mu=6\ {\rm GeV}$, $\sqrt{s_{0}}=6.38\ {\rm GeV}$; the
Borel parameters $T^{2}\in[30,70]\ {\rm GeV}^{2}$. 
\end{itemize}
Our main results are shown in Fig.~\ref{fig:theta_c_and_theta_b}, and the corresponding central values and error estimates are:
\begin{itemize}
\item $\theta_{c}=(1.3\pm0.1)\degree$ from the first sum rule, and $\theta_{c}=(2.0\pm0.8)\degree$
from the second sum rule;
\item $\theta_{b}=(0.31\pm0.03)\degree$ from the first sum rule, and $\theta_{b}=(0.32\pm0.02)\degree$
from the second sum rule. 
\end{itemize}
Here the first and second sum rules respectively refer to those
from the coefficients of $\slashed p$ and constant terms, since
all the $\Pi_{ij}$, at the QCD level, can be computed like 
\begin{equation}
\Pi^{{\rm QCD}}(p)=A(p^{2})\slashed p+B(p^{2}).\label{eq:Pi_QCD}
\end{equation}

In Table~\ref{Tab:comparison}, we compare our results with others
in the literature. It can be seen that, our result for $\theta_{c}$
is consistent with that of Lattice QCD in Ref.~\cite{Liu:2023feb} if the uncertainty is taken into account. 

\begin{table}
\caption{Comparison with other results in the literature (in units of degree). These theoretical predictions respectively come from
QCD sum rules (QCDSR), heavy quark effective theory (HQET), Lattice QCD (LQCD), and quark model (QM).}
\label{Tab:comparison} 
\begin{tabular}{c|c|c|c|c|c|c|c}
\hline 
$\theta_{Q}$ & This work & QCDSR~\cite{Aliev:2010ra} & HQET~\cite{Matsui:2020wcc} & LQCD~\cite{Liu:2023feb} & QM~\cite{Franklin:1996ve} & QM~\cite{Franklin:1981rc} & HQET~\cite{Ito:1996mr} \tabularnewline
\hline 
$\theta_{c}$ & $1.2\sim2.8$ & $5.5\pm1.8$ & $\pm8.12\pm0.80$ & $1.2\pm0.1$ & $3.8$ & $3.8$ & $14\pm14$\tabularnewline
\hline 
$\theta_{b}$ & $0.28\sim0.34$ & $6.4\pm1.8$ & $\pm4.51\pm0.79$ & -- & -- & $1.0$ & --\tabularnewline
\hline 
\end{tabular}
\end{table}

\section{Conclusions and discussions}

\begin{figure}[!]
\includegraphics[width=0.6\columnwidth]{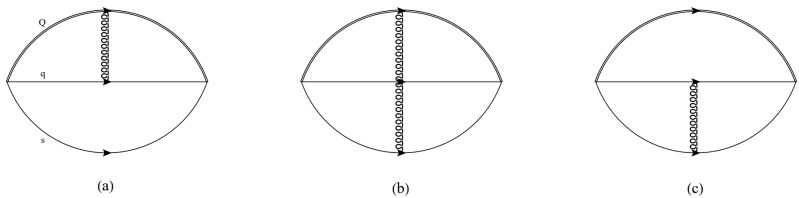}
\caption{These radiative corrections may play an important rule in $\Xi_{Q}-\Xi_{Q}^{\prime}$ mixing.}
\label{fig:radiative_corrections}
\end{figure}

There is a tension between the recent Belle's measurement
and Lattice QCD calculation for the branching ratio of semileptonic
decay ${\cal B}(\Xi_{c}^{0}\to\Xi^{-}e^{+}\nu_{e})$. Some people
proposed that it is possible to resolve this puzzle by considering
the $\Xi_{Q}-\Xi_{Q}^{\prime}$ mixing. Following this suggestion,
we investigate the $\Xi_{Q}-\Xi_{Q}^{\prime}$ mixing using
QCD sum rules in this work. Contributions from up to dimension-6 four-quark
operators are considered. However, it turns out that only dimension-4
and dimension-5 operators contribute, which reveals the non-perturbative
nature of mixing. Especially we notice that only the diagrams with
the two light quarks participating in gluon exchange contribute to
the mixing. Contributions from three-gluon condensate, and
radiative corrections in Fig.~\ref{fig:radiative_corrections} may be sizable and
deserve further investigation. We leave these more detailed consideration for future works.

Our results show that the mixing angle $\theta_{c}$ is very small,
and is consistent with the most recent Lattice QCD calculation result within
error. Such a small mixing
angle seems unlikely to resolve the tension between experimental
measurement and Lattice QCD calculation for the
semileptonic decay $\Xi_{c}^{0}\to\Xi^{-} e^{+}\nu_{e}$.  
We have to draw the conclusion that the tension is still there. 

Finally, it is worth pointing out that Ref.~\cite{Xing:2022phq} recently
proposed a method for measuring the mixing angle in experiment,
which is helpful to further clarify the issue of the $\Xi_{Q}-\Xi_{Q}^{\prime}$
mixing.
\section*{Acknowledgements}

The authors are grateful to Profs.~Yue-Long Shen, Wei Wang, Zhi-Gang Wang,  and Drs.~Hang Liu, Zhi-Peng Xing for valuable discussions. This work is supported in part by National Natural Science Foundation of China under Grant No.~12065020. 

\appendix
\section{Analytical Results}

In this appendix, we present the calculation results of the correlation functions $\Pi_{00,11,01}$ at the QCD level.
Some notes are given below.
\begin{itemize}
\item All non-zero results in Fig. \ref{fig:dim03456} are shown in this
appendix. The spectral densities $\rho^{A}$ and $\rho^{B}$
are shown together. 
\item $m_{1}=m_{Q}$ , $m_{2}=m_{q}$, and $m_{3}=m_{s}$, and $m_{2}$
has been taken to be zero. Because we have defined $m_{23}^{2}\equiv k_{23}^{2}\equiv(k_{2}+k_{3})^{2}$
with $k_{2,3}$ respectively the momenta of the light quark $q$ and
the strange quark, numeric subscripts are preferable. 
\item The $m_{23}^{2}$ appearing in the spectral densities of perturbative diagrams (dimension-0)
and gluon condensate diagrams (dimension-4) should be integrated out. 
\item $\text{m1s}$ is the $m_{1}^{2}$ that appears on the denominator of the propagator of quark 1.
Similar for $\text{m2s}$ and $\text{m3s}$. 
\end{itemize}

\subsection{Results of $\rho_{00}$}

\begin{align}
\rho_{00}^{{\rm dim-0}}=\frac{3}{32\pi^{6}}\Bigg\{ & \frac{m_{1}^{2}-m_{23}^{2}+s}{6m_{23}^{4}s^{2}}\Big[2\left(m_{23}^{2}-m_{3}^{2}\right){}^{2}m_{23}^{2}s\nonumber \\
 & +6m_{3}\left(m_{23}^{2}-m_{3}^{2}\right)m_{23}^{2}\sqrt{s}\left(-m_{1}^{2}+m_{23}^{2}+s\right)\nonumber \\
 & +\left(-2m_{3}^{4}+m_{23}^{2}m_{3}^{2}+m_{23}^{4}\right)\left(-m_{1}^{2}+m_{23}^{2}+s\right){}^{2}\Big],\nonumber \\
 & \frac{m_{1}}{3m_{23}^{4}s}\Big[2\left(m_{23}^{2}-m_{3}^{2}\right){}^{2}m_{23}^{2}s\nonumber \\
 & +6m_{3}\left(m_{23}^{2}-m_{3}^{2}\right)m_{23}^{2}\sqrt{s}\left(-m_{1}^{2}+m_{23}^{2}+s\right)\nonumber \\
 & +\left(-2m_{3}^{4}+m_{23}^{2}m_{3}^{2}+m_{23}^{4}\right)\left(-m_{1}^{2}+m_{23}^{2}+s\right){}^{2}\Big]\Bigg\}\nonumber \\
\times & \frac{\pi\sqrt{\lambda(m_{23}^{2},0,m_{3}^{2})}}{2m_{23}^{2}}\frac{\pi\sqrt{\lambda(s,m_{1}^{2},m_{23}^{2})}}{2s},
\end{align}

\begin{align}
\rho_{00}^{{\rm dim-3(a)}}=-\frac{\langle\bar{q}q\rangle}{16\pi^{3}}\Bigg\{ & \frac{2\left(m_{1}^{2}-m_{3}^{2}+s\right)\left(\left(m_{3}+\sqrt{s}\right){}^{2}-m_{1}^{2}\right)}{s^{3/2}},\nonumber \\
 & \frac{4m_{1}\left(\left(m_{3}+\sqrt{s}\right){}^{2}-m_{1}^{2}\right)}{\sqrt{s}}\Bigg\}\times\frac{\pi\sqrt{\lambda(s,m_{1}^{2},m_{3}^{2})}}{2s},
\end{align}

\begin{equation}
\rho_{00}^{{\rm dim-3(b)}}=-\frac{\langle\bar{s}s\rangle}{16\pi^{3}}\left\{ \frac{2\left(s^{2}-m_{1}^{4}\right)}{s^{3/2}},\frac{4m_{1}\left(s-m_{1}^{2}\right)}{\sqrt{s}}\right\} \times\frac{\pi\sqrt{\lambda(s,m_{1}^{2},0)}}{2s},
\end{equation}

\begin{align}
\rho_{00}^{{\rm dim-4(c)}}= & \left(-\frac{\text{\ensuremath{\langle g_{s}^{2}G^{2}\rangle}}}{24576\pi^{6}}\right)\frac{\partial}{\partial\text{m2s}}\frac{\partial}{\partial\text{m3s}}\Bigg\{\frac{16\left(m_{1}^{2}-m_{23}^{2}+s\right)}{s^{2}}\Big[-\frac{1}{6m_{23}^{4}}\nonumber \\
 & \times\Big(\left(-m_{1}^{2}+m_{23}^{2}+s\right){}^{2}\left(m_{23}^{4}+m_{23}^{2}(\text{m2s}+\text{m3s})-2(\text{m2s}-\text{m3s})^{2}\right)\nonumber \\
 & \quad+2m_{23}^{2}s\left(m_{23}^{4}-2m_{23}^{2}(\text{m2s}+\text{m3s})+(\text{m2s}-\text{m3s})^{2}\right)\Big)\nonumber \\
 & -\frac{3m_{3}\sqrt{s}\left(-m_{1}^{2}+m_{23}^{2}+s\right)(m_{23}^{2}+\text{m2s}-\text{m3s})}{m_{23}^{2}}\nonumber \\
 & -4s(m_{23}^{2}-\text{m2s}-\text{m3s})\Big],\nonumber \\
 & 32m_{1}\Big[-\frac{1}{6m_{23}^{4}s}\nonumber \\
 & \times\Big(\left(-m_{1}^{2}+m_{23}^{2}+s\right){}^{2}\left(m_{23}^{2}(\text{m2s}+\text{m3s})+m_{23}^{4}-2(\text{m2s}-\text{m3s})^{2}\right)\nonumber \\
 & \quad+2m_{23}^{2}s\left(m_{23}^{4}-2m_{23}^{2}(\text{m2s}+\text{m3s})+(\text{m2s}-\text{m3s})^{2}\right)\Big)\nonumber \\
 & -\frac{3m_{3}\left(-m_{1}^{2}+m_{23}^{2}+s\right)\left(m_{23}^{2}+\text{m2s}-\text{m3s}\right)}{m_{23}^{2}\sqrt{s}}\nonumber \\
 & -4\left(m_{23}^{2}-\text{m2s}-\text{m3s}\right)\Big]\Bigg\}\nonumber \\
\times & \frac{\pi\sqrt{\lambda(m_{23}^{2},\text{m2s},\text{m3s})}}{2m_{23}^{2}}\frac{\pi\sqrt{\lambda(s,m_{1}^{2},m_{23}^{2})}}{2s},
\end{align}

\begin{align}
\rho_{00}^{{\rm dim-4(d)}}= & \frac{\langle g_{s}^{2}G^{2}\rangle}{128\pi^{6}}\frac{1}{6}\frac{\partial^{3}}{\partial\text{m1s}^{3}}\Bigg\{\frac{m_{1}^{2}\left(-m_{23}^{2}+\text{m1s}+s\right)}{6m_{23}^{4}s^{2}}\nonumber \\
 & \times\Big[6m_{3}\left(m_{23}^{2}-m_{3}^{2}\right)m_{23}^{2}\sqrt{s}\left(m_{23}^{2}-\text{m1s}+s\right)\nonumber \\
 & \quad+\left(-2m_{3}^{4}+m_{23}^{2}m_{3}^{2}+m_{23}^{4}\right)\left(m_{23}^{2}-\text{m1s}+s\right){}^{2}\nonumber \\
 & \quad+2\left(m_{23}^{2}-m_{3}^{2}\right){}^{2}m_{23}^{2}s\Big],\nonumber \\
 & \frac{m_{1}\text{m1s}}{3m_{23}^{4}s}\nonumber \\
 & \times\Big[6m_{3}\left(m_{23}^{2}-m_{3}^{2}\right)m_{23}^{2}\sqrt{s}\left(m_{23}^{2}-\text{m1s}+s\right)\nonumber \\
 & \quad+\left(-2m_{3}^{4}+m_{23}^{2}m_{3}^{2}+m_{23}^{4}\right)\left(m_{23}^{2}-\text{m1s}+s\right){}^{2}\nonumber \\
 & \quad+2\left(m_{23}^{2}-m_{3}^{2}\right){}^{2}m_{23}^{2}s\Big]\Bigg\}\nonumber \\
\times & \frac{\pi\sqrt{\lambda(m_{23}^{2},0,m_{3}^{2})}}{2m_{23}^{2}}\frac{\pi\sqrt{\lambda(s,\text{m1s},m_{23}^{2})}}{2s},
\end{align}

\begin{align}
\rho_{00}^{{\rm dim-5(b)}}= & \left(-\frac{\langle\bar{q}g_{s}\sigma Gq\rangle}{1536\pi^{3}}\right)\frac{\partial}{\partial\text{m3s}}\Bigg\{-\frac{24\left(m_{1}^{2}-\text{m3s}+s\right)\left(2m_{3}\sqrt{s}-m_{1}^{2}+\text{m3s}+s\right)}{s^{3/2}},\nonumber \\
 & -4m_{1}\left(\frac{12\left(-m_{1}^{2}+\text{m3s}+s\right)}{\sqrt{s}}+24m_{3}\right)\Bigg\}\frac{\pi\sqrt{\lambda(s,m_{1}^{2},\text{m3s})}}{2s},
\end{align}

\begin{align}
\rho_{00}^{{\rm dim-5(d)}}=\left(-\frac{\langle\bar{s}g_{s}\sigma Gs\rangle}{1536\pi^{3}}\right)\frac{\partial}{\partial\text{m2s}}\Bigg\{ & -\frac{24\left(-m_{1}^{2}+\text{m2s}+s\right)\left(m_{1}^{2}-\text{m2s}+s\right)}{s^{3/2}},\nonumber \\
 & -\frac{48m_{1}\left(-m_{1}^{2}+\text{m2s}+s\right)}{\sqrt{s}}\Bigg\}\frac{\pi\sqrt{\lambda(s,m_{1}^{2},\text{m2s})}}{2s},
\end{align}

\begin{equation}
\rho_{00}^{{\rm dim-6}}=\left(\frac{\langle\bar{q}q\rangle\langle\bar{s}s\rangle}{24}\right)\{8,8m_{1}\}\delta(s-m_{1}^{2}).
\end{equation}

\subsection{Results of $\rho_{11}$}

\begin{align}
\rho_{11}^{{\rm dim-0}} & =\rho_{00}^{{\rm dim-0}},\\
\rho_{11}^{{\rm dim-3(a)}} & =\rho_{00}^{{\rm dim-3(a)}},\\
\rho_{11}^{{\rm dim-3(b)}} & =\rho_{00}^{{\rm dim-3(b)}},
\end{align}

\begin{align}
\rho_{11}^{{\rm dim-4}(a)}= & \left(-\frac{\langle g_{s}^{2}G^{2}\rangle}{24576\pi^{6}}\right)\frac{\partial}{\partial\text{m1s}}\frac{\partial}{\partial\text{m2s}}\Bigg\{-\frac{16}{9m_{23}^{4}s}\nonumber \\
 & \times\Big[-24m_{3}m_{23}^{2}\sqrt{s}\left(-m_{3}^{2}+m_{23}^{2}+\text{m2s}\right)\left(m_{23}^{2}+\text{m1s}-s\right)\nonumber \\
 & +\frac{6m_{3}m_{23}^{2}\left(-m_{3}^{2}+m_{23}^{2}+\text{m2s}\right)\left(m_{23}^{2}-\text{m1s}+s\right)\left(-m_{23}^{2}+\text{m1s}+s\right)}{\sqrt{s}}\nonumber \\
 & +4\Big(m_{23}^{2}\left(-2m_{3}^{2}\left(m_{23}^{2}+\text{m2s}\right)+\left(m_{23}^{2}-\text{m2s}\right){}^{2}+m_{3}^{4}\right)\left(-m_{23}^{2}+\text{m1s}+s\right)\nonumber \\
 & \quad+\left(m_{23}^{2}\left(m_{3}^{2}+\text{m2s}\right)-2\left(\text{m2s}-m_{3}^{2}\right){}^{2}+m_{23}^{4}\right)\nonumber \\
 & \qquad\times\left(-m_{23}^{2}-\text{m1s}+s\right)\left(m_{23}^{2}-\text{m1s}+s\right)\Big)\nonumber \\
 & +\frac{\left(-m_{23}^{2}+\text{m1s}+s\right)}{s}\nonumber \\
 & \times\Big(\left(m_{23}^{2}\left(m_{3}^{2}+\text{m2s}\right)-2\left(\text{m2s}-m_{3}^{2}\right){}^{2}+m_{23}^{4}\right)\left(m_{23}^{2}-\text{m1s}+s\right){}^{2}\nonumber \\
 & \quad+2m_{23}^{2}s\left(-2m_{3}^{2}\left(m_{23}^{2}+\text{m2s}\right)+\left(m_{23}^{2}-\text{m2s}\right){}^{2}+m_{3}^{4}\right)\Big)\Big],\nonumber \\
 & -\frac{32m_{1}}{3m_{23}^{4}s}\nonumber \\
 & \times\Big[\left(m_{23}^{2}\left(m_{3}^{2}+\text{m2s}\right)-2\left(\text{m2s}-m_{3}^{2}\right){}^{2}+m_{23}^{4}\right)\left(m_{23}^{2}-\text{m1s}+s\right){}^{2}\nonumber \\
 & \quad+6m_{3}m_{23}^{2}\sqrt{s}\left(-m_{3}^{2}+m_{23}^{2}+\text{m2s}\right)\left(m_{23}^{2}-\text{m1s}+s\right)\nonumber \\
 & \quad+2m_{23}^{2}s\left(-2m_{3}^{2}\left(m_{23}^{2}+\text{m2s}\right)+\left(m_{23}^{2}-\text{m2s}\right){}^{2}+m_{3}^{4}\right)\Big]\Bigg\}\nonumber \\
\times & \frac{\pi\sqrt{\lambda(m_{23}^{2},\text{m2s},m_{3}^{2})}}{2m_{23}^{2}}\frac{\pi\sqrt{\lambda(s,\text{m1s},m_{23}^{2})}}{2s},
\end{align}

\begin{align}
\rho_{11}^{{\rm dim-4(b)}}= & \left(-\frac{\langle g_{s}^{2}G^{2}\rangle}{24576\pi^{6}}\right)\frac{\partial}{\partial\text{m1s}}\frac{\partial}{\partial\text{m3s}}\Bigg\{\frac{16}{9m_{23}^{4}s}\nonumber \\
 & \times\Big[\frac{2\left(m_{23}^{2}\text{m3s}+m_{23}^{4}-2\text{m3s}^{2}\right)\left(-m_{23}^{2}+\text{m1s}+s\right)\left(m_{23}^{2}-\text{m1s}+s\right){}^{2}}{s}\nonumber \\
 & \quad+4\left(m_{23}^{2}\text{m3s}+m_{23}^{4}-2\text{m3s}^{2}\right)\left(m_{23}^{2}+\text{m1s}-s\right)\left(m_{23}^{2}-\text{m1s}+s\right)\nonumber \\
 & \quad-\frac{3\left(-m_{23}^{2}+\text{m1s}+s\right)}{s}\Big(\left(m_{23}^{2}\text{m3s}+m_{23}^{4}-2\text{m3s}^{2}\right)\left(m_{23}^{2}-\text{m1s}+s\right){}^{2}\nonumber \\
 & \qquad+2m_{23}^{2}s\left(m_{23}^{2}-\text{m3s}\right){}^{2}\Big)\nonumber \\
 & \quad-\frac{18m_{3}m_{23}^{2}\left(m_{23}^{2}-\text{m3s}\right)\left(-m_{23}^{2}+\text{m1s}+s\right)\left(m_{23}^{2}-\text{m1s}+s\right)}{\sqrt{s}}\Big],\nonumber \\
 & -\frac{32m_{1}\left(m_{23}^{2}-\text{m3s}\right)}{3m_{23}^{4}s}\nonumber \\
 & \times\Big[\text{m1s}^{2}\left(m_{23}^{2}+2\text{m3s}\right)-2\text{m1s}\left(m_{23}^{2}+2\text{m3s}\right)\left(m_{23}^{2}+s\right)\nonumber \\
 & \quad+6m_{3}m_{23}^{2}\sqrt{s}\left(m_{23}^{2}-\text{m1s}+s\right)+2m_{23}^{2}\text{m3s}s+2m_{23}^{4}\text{m3s}\nonumber \\
 & \quad+m_{23}^{2}s^{2}+4m_{23}^{4}s+m_{23}^{6}+2\text{m3s}s^{2}\Big]\Bigg\}\nonumber \\
\times & \frac{\pi\sqrt{\lambda(m_{23}^{2},0,\text{m3s})}}{2m_{23}^{2}}\frac{\pi\sqrt{\lambda(s,\text{m1s},m_{23}^{2})}}{2s},
\end{align}

\begin{align}
\rho_{11}^{{\rm dim-4(c)}} & =-\frac{1}{3}\rho_{00}^{{\rm dim-4(c)}},
\end{align}

\begin{align}
\rho_{11}^{{\rm dim-4(d)}} & =\rho_{00}^{{\rm dim-4(d)}},
\end{align}

\begin{align}
\rho_{11}^{{\rm dim-5(a)}}= & \left(-\frac{\langle\bar{q}g_{s}\sigma Gq\rangle}{1536\pi^{3}}\right)\frac{\partial}{\partial\text{m1s}}\Bigg\{-\frac{16\left(-m_{3}^{2}+\text{m1s}+s\right)\left(2m_{3}\sqrt{s}+m_{3}^{2}-\text{m1s}+s\right)}{s^{3/2}},\nonumber \\
 & -\frac{32m_{1}\left(2m_{3}\sqrt{s}+m_{3}^{2}-\text{m1s}+s\right)}{\sqrt{s}}\Bigg\}\frac{\pi\sqrt{\lambda(s,\text{m1s},m_{3}^{2})}}{2s},
\end{align}

\begin{equation}
\rho_{11}^{{\rm dim-5(b)}}=-\frac{1}{3}\rho_{00}^{{\rm dim-5(b)}},
\end{equation}

\begin{align}
\rho_{11}^{{\rm dim-5(c)}}=\left(-\frac{\langle\bar{s}g_{s}\sigma Gs\rangle}{1536\pi^{3}}\right)\frac{\partial}{\partial\text{m1s}} & \Bigg\{\frac{16(\text{m1s}-s)(\text{m1s}+s)}{s^{3/2}},\frac{32m_{1}(\text{m1s}-s)}{\sqrt{s}}\Bigg\}\nonumber \\
 & \times\frac{\pi\sqrt{\lambda(s,\text{m1s},0)}}{2s},
\end{align}
\begin{equation}
\rho_{11}^{{\rm dim-5(d)}}=-\frac{1}{3}\rho_{00}^{{\rm dim-5(d)}},
\end{equation}

\begin{equation}
\rho_{11}^{{\rm dim-6}}=\rho_{00}^{{\rm dim-6}}.
\end{equation}

\subsection{Results of $\rho_{01}$}

\begin{align}
\rho_{01}^{{\rm dim-5(a)}}= & \left(-\frac{\langle\bar{q}g_{s}\sigma Gq\rangle}{1536\pi^{3}}\right)\frac{\partial}{\partial\text{m1s}}\Bigg\{-\frac{16\sqrt{3}m_{1}\left(2m_{3}\sqrt{s}+m_{3}^{2}-\text{m1s}+s\right)}{s},\nonumber \\
 & -\frac{8\sqrt{3}\left(-m_{3}^{2}+\text{m1s}+s\right)\left(2m_{3}\sqrt{s}+m_{3}^{2}-\text{m1s}+s\right)}{s}\Bigg\}\frac{\pi\sqrt{\lambda(s,\text{m1s},m_{3}^{2})}}{2s},
\end{align}

\begin{align}
\rho_{01}^{{\rm dim-5(c)}}=\left(-\frac{\langle\bar{s}g_{s}\sigma Gs\rangle}{1536\pi^{3}}\right)\frac{\partial}{\partial\text{m1s}} & \Bigg\{\frac{16\sqrt{3}m_{1}(s-\text{m1s})}{s},\frac{8\sqrt{3}(s-\text{m1s})(\text{m1s}+s)}{s}\Bigg\}\nonumber \\
 & \times\frac{\pi\sqrt{\lambda(s,\text{m1s},0)}}{2s},
\end{align}

\begin{align}
\rho_{01}^{{\rm dim-4(a)}}= & \left(-\frac{\text{\ensuremath{\langle g_{s}^{2}G^{2}\rangle}}}{24576\pi^{6}}\right)\frac{\partial}{\partial\text{m1s}}\frac{\partial}{\partial\text{m2s}}\Bigg\{-\frac{16m_{1}}{\sqrt{3}m_{23}^{4}s}\nonumber \\
 & \times\Big[6m_{3}m_{23}^{2}\left(-m_{3}^{2}+m_{23}^{2}+\text{m2s}\right)\left(m_{23}^{2}-\text{m1s}+s\right)\nonumber \\
 & \quad+\frac{1}{\sqrt{s}}\Big(\left(m_{23}^{2}\left(m_{3}^{2}+\text{m2s}\right)-2\left(\text{m2s}-m_{3}^{2}\right){}^{2}+m_{23}^{4}\right)\left(m_{23}^{2}-\text{m1s}+s\right){}^{2}\nonumber \\
 & \qquad+2m_{23}^{2}s\left(-2m_{3}^{2}\left(m_{23}^{2}+\text{m2s}\right)+\left(m_{23}^{2}-\text{m2s }\right){}^{2}+m_{3}^{4}\right)\Big)\Big],\nonumber \\
 & -\frac{8}{3\sqrt{3}m_{23}^{4}}\Big[\frac{\left(-m_{23}^{2}+\text{m1s}+s\right)}{s^{3/2}}\nonumber \\
 & \quad\times\Big(\left(m_{23}^{2}\left(m_{3}^{2}+\text{m2s}\right)-2\left(\text{m2s}-m_{3}^{2}\right){}^{2}+m_{23}^{4}\right)\left(m_{23}^{2}-\text{m1s}+s\right){}^{2}\nonumber \\
 & \qquad+2m_{23}^{2}s\left(-2m_{3}^{2}\left(m_{23}^{2}+\text{m2s}\right)+\left(m_{23}^{2}-\text{m2s}\right){}^{2}+m_{3}^{4}\right)\Big)\nonumber \\
 & \quad-24m_{3}m_{23}^{2}\left(-m_{3}^{2}+m_{23}^{2}+\text{m2s}\right)\left(m_{23}^{2}+\text{m1s}-s\right)\nonumber \\
 & \quad+\frac{6m_{3}m_{23}^{2}\left(-m_{3}^{2}+m_{23}^{2}+\text{m2s}\right)\left(m_{23}^{2}-\text{m1s}+s\right)\left(-m_{23}^{2}+\text{m1s}+s\right)}{s}\nonumber \\
 & \quad+\frac{4}{\sqrt{s}}\Big(m_{23}^{2}\left(-2m_{3}^{2}\left(m_{23}^{2}+\text{m2s}\right)+\left(m_{23}^{2}-\text{m2s}\right){}^{2}+m_{3}^{4}\right)\left(-m_{23}^{2}+\text{m1s}+s\right)\nonumber \\
 & \qquad+\left(m_{23}^{2}\left(m_{3}^{2}+\text{m2s}\right)-2\left(\text{m2s}-m_{3}^{2}\right){}^{2}+m_{23}^{4}\right)\nonumber \\
 & \qquad\quad\times\left(-m_{23}^{2}-\text{m1s}+s\right)\left(m_{23}^{2}-\text{m1s}+s\right)\Big)\Big]\Bigg\}\nonumber \\
\times & \frac{\pi\sqrt{\lambda(m_{23}^{2},\text{m2s},m_{3}^{2})}}{2m_{23}^{2}}\frac{\pi\sqrt{\lambda(s,\text{m1s},m_{23}^{2})}}{2s},
\end{align}

\begin{align}
\rho_{01}^{{\rm dim-4(b)}}= & \left(-\frac{\langle g_{s}^{2}G^{2}\rangle}{24576\pi^{6}}\right)\frac{\partial}{\partial\text{m1s}}\frac{\partial}{\partial\text{m3s}}\Bigg\{\frac{16m_{1}}{\sqrt{3}m_{23}^{4}s}\nonumber \\
 & \times\Big[\frac{\left(m_{23}^{2}\text{m3s}+m_{23}^{4}-2\text{m3s}^{2}\right)\left(m_{23}^{2}-\text{m1s}+s\right){}^{2}+2m_{23}^{2}s\left(m_{23}^{2}-\text{m3s}\right){}^{2}}{\sqrt{s}}\nonumber \\
 & \quad+6m_{3}m_{23}^{2}\left(m_{23}^{2}-\text{m3s}\right)\left(m_{23}^{2}-\text{m1s}+s\right)\Big],\nonumber \\
 & \frac{8}{3\sqrt{3}m_{23}^{4}s^{3/2}}\nonumber \\
 & \times\Big[4s\Big(m_{23}^{2}\left(m_{23}^{2}-\text{m3s}\right){}^{2}\left(-m_{23}^{2}+\text{m1s}+s\right)\nonumber \\
 & \qquad-\left(m_{23}^{2}\text{m3s}+m_{23}^{4}-2\text{m3s}^{2}\right)\left(m_{23}^{2}+\text{m1s}-s\right)\left(m_{23}^{2}-\text{m1s}+s\right)\Big)\nonumber \\
 & \quad+\left(-m_{23}^{2}+\text{m1s}+s\right)\Big(\left(m_{23}^{2}\text{m3s}+m_{23}^{4}-2\text{m3s}^{2}\right)\left(m_{23}^{2}-\text{m1s}+s\right){}^{2}\nonumber \\
 & \qquad+2m_{23}^{2}s\left(m_{23}^{2}-\text{m3s}\right){}^{2}\Big)\nonumber \\
 & \quad+18m_{3}m_{23}^{2}\sqrt{s}\left(m_{23}^{2}-\text{m3s}\right)\left(-m_{23}^{2}+\text{m1s}+s\right)\left(m_{23}^{2}-\text{m1s}+s\right)\Big]\Bigg\}\nonumber \\
\times & \frac{\pi\sqrt{\lambda(m_{23}^{2},0,\text{m3s})}}{2m_{23}^{2}}\frac{\pi\sqrt{\lambda(s,\text{m1s},m_{23}^{2})}}{2s}.
\end{align}

\end{document}